\documentclass[journal]{IEEEtran}
\usepackage{xcolor,soul}

\usepackage{cite}
\usepackage{color}
\usepackage[mathlines]{lineno}

\usepackage[utf8]{inputenc}
\usepackage[T1]{fontenc}

\usepackage{csquotes}
\usepackage[font={small,it}]{caption}
\usepackage{amsmath}
\usepackage{upgreek}
\usepackage{wasysym}
\usepackage{amsfonts}
\usepackage{graphicx}
\usepackage{hyperref}

\begin{document}
\twocolumn[
\begin{@twocolumnfalse}
{\Huge{IEEE Copyright Notice}}\\
\\
\large Copyright (c) 2018 IEEE\\
Personal use of this material is permitted. Permission from IEEE must be obtained for all other
uses, in any current or future media, including reprinting/republishing this material for advertising
or promotional purposes, creating new collective works, for resale or redistribution to servers or lists,
or reuse of any copyrighted component of this work in other works.
\\
\\
To be published in IEEE Geoscience and Remote Sensing Letters.\\
\\
Digital Object Identifier 10.1109/LGRS.2018.2812770
\end{@twocolumnfalse}
]

\newpage

\title{A New Model for the Distribution of Observable Earthquake Magnitudes and Applications to $b$-value Estimation}
\author{Jesper Martinsson and Adam Jonsson
\thanks{Manuscript received November 12, 2017; accepted February 28, 2018. Date of publication X X, 2018; date of current version X X, 2018}%
\thanks{J. Martinsson and A. Jonsson are with the Department of Engineering Sciences and Mathematics, Lule{\aa} University of Technology, 97187 Lule{\aa}, Sweden (e-mail: jesper.martinsson@ltu.se; adam.jonsson@ltu.se).}%
\thanks{J. Martinsson is also with the Department of Mining Technology, Luossavaara-Kiirunavaara AB (LKAB), 98381 Malmberget, Sweden (e-mail: jesper.martinsson@lkab.com).}}

\maketitle

\begin{abstract} 
The $b$-value in the Gutenberg-Richter (GR) law contains information that is essential for evaluating  earthquake hazard and predicting the occurrence of large earthquakes. Estimates of $b$ are often based on seismic events whose magnitude exceed a certain threshold, the so called magnitude of completeness. Such estimates are sensitive to the choice of threshold  and often ignore a substantial portion of available data. We present a general model for the distribution of observable earthquake magnitudes and an estimation procedure that takes all measurements into account. The model is obtained by  generalizing previous probabilistic descriptions of sensor network limitations  and by using a generalization of the GR law. We show that our model is flexible enough to handle spatio-temporal variations in the seismic  environment and captures valuable information about sensor network coverage. We also show that the model leads to significantly improved $b$-value estimates compared with established methods relying on the magnitude of completeness. 
\end{abstract}


\section{Introduction}
Estimating the $b$-value in the Gutenberg-Richter (GR) law is important in analyzing seismicity and essential for predicting the occurrence of extreme events \cite{SC17}.  
The low detection rate for seismic events with small magnitudes presents a well known challenge for efficient estimation: the observed magnitudes do not follow the GR law. This problem is often handled by discarding all events with magnitudes below a certain threshold, commonly referred to as the magnitude of completeness and denoted $m_\mathrm{c}$ here. The determination of this threshold by visual inspection or automatic routines  then becomes crucial for estimating the $b$-value. On the one hand, since an unbiased estimate requires a detection rate of 100\% above the threshold, a too small value of $m_\mathrm{c}$ introduces bias. On the other hand, although the bias vanishes    as $m_\mathrm{c}$ increases, a high threshold will  increase uncertainty as more data is discarded. These difficulties account for the diversity of existing techniques and algorithms  for determining $m_\mathbf{c}$ \cite{RS89, WW00, MW12, CG02, Mar03, WW05, Amo07, SW08, SMM10}. 

Since real sensor networks will fail to guarantee a 100\% detection rate, it may be argued that their limitations are more appropriately modeled probabilistically. This article generalizes previous work \cite{Rin75,OK93,OK06,Iwa08} in this direction by introducing a class of detection probability functions that is flexible enough to model inhomogeneous conditions arising from spatio-temporal variations---in the environment, in the detecability of events, and in the sensitivity of the sensor network (e.g., network expansions). Our aim is to investigate whether a more flexible model results in better estimates when information is used from all available measurements.  
 
\section{The Detection Probability}
We model the sensitivity of a sensor network by its detection probability, which we define as the probability that an event of magnitude $m$ is detected. A commonly-used criterion for an event to be processed and registered is that the event is detected by at least $k$ sensors, for some specified value of  $k$. The detection probability  then becomes the probability that an event of magnitude $m$ is detected by at least $k$ sensors. We write this probability as $P(R \leq r_m)$, where $R$ denotes the hypocenter distance from the event to the $k$-nearest sensor and where $r_m$ is the maximum distance from which an event with magnitude $m$ can be detected. 

In stationary networks, the distribution of $R$  is appropriately modeled by the generalized gamma distribution \cite{Hae05}. To describe realistic sensor networks, allowing spatio-temporal variations, we hypothesize that non-stationary environments can be represented as mixtures of stationary environments with different characteristics. More precisely, we adopt a mixture model where the probability density function (PDF) of $R$ is a convex combination $\sum_{i=0}^{I-1} \phi_i f_i$, where $f_i$ is the PDF of the log-normal distribution $\ln N(\mu_i,\sigma_i)$. We choose log-normal distributions for three reasons: (i) the log-normal distribution is a special case of the generalized gamma distribution, (ii) its shape resembles the generalized gamma with the parameter values and dimensions of interest here, and (iii) it yields mathematical tractability in the subsequent analysis.

Under the stated assumptions, the detection probability becomes
\begin{linenomath*}
\begin{align}\label{eq: first sensitivity function}
\hspace{-0.25cm}  P(R\leq r_m)=\sum_{i=0}^{I-1} \phi_i\left(\frac{1}{2}+\frac{1}{2}\mathrm{erf}\left(\frac{\ln(r_m)-\mu_i}{\sqrt{2}\sigma_i}\right)\right).
\end{align}
\end{linenomath*}
To develop the expression in (\ref{eq: first sensitivity function}), we need assumptions on how $r_m$ depends on $m$. Denoting the amplitude of a wave measured by a sensor at distance $r$ from the event by $A(r)$, a simple model is given by (see e.g. \cite{Ric35})
\begin{linenomath*}
\begin{align}\label{eq: equation for A}
A(r)=\exp(\alpha_0+\alpha_1 m) r^{-\alpha_2},
\end{align}
\end{linenomath*}
where the parameters $\alpha_0$, $\alpha_1$ and $\alpha_2$ describe the amplitude, magnitude and wave field relationships, respectively. 

If $A_{\mathrm{min}}$ denotes the amplitude required for a sensor to detect the waveform, then \begin{linenomath*}
$$r_m=\exp((\alpha_0+\alpha_1 m)/\alpha_2) A_{\mathrm{min}}^{-1/\alpha_2}$$\end{linenomath*} is the corresponding maximum detectable distance of an event with magnitude $m$. Inserting $r_m$ in (\ref{eq: first sensitivity function}) gives
\begin{linenomath*}
\begin{align}\label{eq:P(R<=r_m)}
P(&R\leq r_m)=\sum_{i=0}^{I-1} \phi_i\left(\frac{1}{2}+\frac{1}{2}\mathrm{erf}\left(\frac{m-\tilde{\mu}_i}{\sqrt{2}\tilde{\sigma}_i}\right)\right),
\end{align}
\end{linenomath*}
where the relationships 
\begin{linenomath*}
\begin{align}
\tilde{\mu}_i &= (\alpha_2\mu_i + \ln(A_{\mathrm{min}})-\alpha_0)\alpha_1^{-1},\label{eq:relationship1}\\
\tilde{\sigma}_i&=\sigma_i\alpha_2\alpha_1^{-1},\label{eq:relationship2}
\end{align}
\end{linenomath*}
apply for conditions where (\ref{eq: equation for A}) holds. The convex combination  (\ref{eq:P(R<=r_m)}) of terms with different characteristics will capture the sensor network's limitations and sensitivity for the exported data regardless of whether (\ref{eq: equation for A})  holds or not. For events that do not follow  (\ref{eq: equation for A}),  the model (\ref{eq:P(R<=r_m)}) is still applicable, but its parameters cannot be expressed in terms of the parameters in (\ref{eq: equation for A}). In this case,  (\ref{eq:relationship1})  and (\ref{eq:relationship2}) do not apply.

\section{The Observed Magnitude Distribution}
The PDF of the observed magnitudes can be written $f_M(m)=P(R\leq r_m)g(m)$, where $g$ gives the relationship between magnitude and frequency. Under stationary conditions, this  relationship is appropriately modeled by the GR law. To allow for spatio-temporal variations in real data \cite{Imo91,PAD11,Zha17}, we generalize the GR law, writing $g$ as a convex combination
\begin{linenomath*}
\begin{align}\label{eq:g(m)}
g(m) = \sum_{j=0}^{J-1}\omega_j\lambda_j\exp(-\lambda_j m), 
\end{align}
\end{linenomath*}
of $J$ possible characteristics with  weights $\omega_j$ and $b$-values $b_j=\lambda_j/\ln(10)$. (Here and below, the term  $b$-value refers to either $b$ or $\lambda$.) The PDF of the resulting observable magnitude distribution is then given by
\begin{linenomath*}
\begin{align}
 \nonumber f_M(m) = \sum_{i=0}^{I-1}& \phi_i\left(\frac{1}{2}+\frac{1}{2}\mathrm{erf}\left(\frac{m-\tilde{\mu}_i}{\sqrt{2}\tilde{\sigma}_i}\right)\right)\\
\times & \sum_{j=0}^{J-1}\omega_j\lambda_j\exp(-\lambda_j m) c^{-1}_{ij}, \label{eq:f_M(m)}
\end{align}
\end{linenomath*}
where $c_{ij}=\exp(\lambda_j^2 \tilde{\sigma}_i^2/2-\tilde{\mu}_i\lambda_j)$
is the normalizing factor to obtain unit probability mass.

The distribution in (\ref{eq:f_M(m)}) can be seen as a generalization of previous work to describe the entire magnitude range; see \cite{Rin75,OK93,OK06,Iwa08}. In fact, the distribution (\ref{eq:f_M(m)}) in its simplest form ($I=1,J=1$) includes the ones studied by these authors. When $\tilde{\sigma}_0\rightarrow 0$, the simplest form can also describe data truncated at $m_\mathrm{c}$ by setting $\tilde{\mu}_0=m_\mathrm{c}$. 

\section{Results and Discussion}
To test the validity of the model, and to investigate whether or not the increased flexibility improves inference when the data have spatio-temporal variations, we  applied the model (\ref{eq:f_M(m)}) to five data sets from three sensor networks: We study three data sets from the Southern California Seismic Network (\href{https://service.scedc.caltech.edu/eq-catalogs/date_mag_loc.php}{SCSN}) with varying sensor network sensitivity \cite{KWH10}. We consider one large data set from the National Research Institute for Earth Science and Disaster Prevention (\href{http://www.fnet.bosai.go.jp/event/search.php?LANG=en}{NIED}), Japan, which has less sensor-network variations \cite{OKHO04}. In addition to tectonic movements, we also include an example of mining induced seismicity in Sweden to demonstrate the importance of generalizing the GR law in (\ref{eq:g(m)}). In each case, the model was fitted to the data sets without any of the pre-processing strategies often recommended to obtain reliable $b$-value estimates. For the SCSN and NIED data sets, we used the default settings for geometrical restrictions, obtaining aggregated data from regions with varying sensor network coverage.

Our model allows us to describe the observed magnitudes to a high degree of accuracy by taking $I$ and $J$ large. We use Maximum Likelihood (ML) to estimate the model parameters and rely on the Bayesian Information Criterion (BIC) \cite{Sch78} to avoid over-parametrisation \cite{SS04}. For model validation, we  assess replicated data from the distribution to detect aspects of the data not captured by the model \cite{GCSR04}. As a discrepancy measure, we consider the replicated uncertainties in each magnitude bin, as this measure reflects our   interest in providing a description of the entire magnitude range. This approach has advantages \cite{GMS96} compared with standard test statistics (e.g. $\chi^2$ or likelihood ratio): it enables us to assess if the information criterion is too restrictive, where discrepancies are located, and if systematic deviations are present.

Figs.~\ref{fig:dataset_1}--\ref{fig:dataset_4} show the fitted models (solid black curve) for our five data sets in (a) linear and (b) logarithmic vertical scales. The dots give the relative frequency of the observed magnitudes. The vertical bars indicate the 2.5 and 97.5 percentiles of 100,000 replicated data sets of equal size from the proposed distribution. Under the assumption that our model gives a valid description of observable magnitudes, we would expect the vertical bars to capture about 95\% of the observed dots, and no systematic patterns should be observed for dots lying outside the vertical bars. The solid red curve gives the ML estimate of the $b$-value \cite{Aki65,Uts65,Ben83} based on events with magnitude $m>m_\mathrm{c}$ and the Goodness-of-Fit Test (GFT) \cite{WW00} to obtain $m_\mathrm{c}$. In Figs.~\ref{fig:dataset_8}--\ref{fig:dataset_6}, the fitted model is also compared with a model (dashed green curve) involving a single term ($I=1,J=1$). The latter coincides with the model in \cite{Rin75,OK93,OK06,Iwa08}.

For the data set in Fig.~\ref{fig:dataset_1}, the ML estimate with $m_\mathrm{c}=1.0$ indicates an underestimation of the threshold and explains the visible difference in $b$-values between the two procedures. A reduction of the detection rate can be observed as high as $m=1.2$, where  magnitudes at 1.2 and below will significantly influence the ML estimate of the $b$-value.

\def\mycap{{\captitle} Observations and fitted probability density in (a) linear and (b) logarithmic vertical scales. The estimated $b$-value for the fitted PDF, $f_M(m)$, is $\caplhat$. The ML-estimate using the threshold  $m_\mathrm{c}=\capmc$ is $\hat{\lambda}_\mathrm{ML}=
\caplmlhat$. The fraction of observations inside the 95\% vertical bars is $\capfrac$. The 95\% CI is $\capCI$.}

\def\mycapalt{{\captitle} Observations and fitted probability density in (a) linear and (b) logarithmic vertical scales. The estimated $b$-value for the fitted PDF, $f_M(m)$, is $\caplhat$. The ML-estimate using the threshold  $m_\mathrm{c}=\capmc$ is $\hat{\lambda}_\mathrm{ML}=
\caplmlhat$. The 95\% vertical bars contain $\capfraca$ out of the $\capfracb$ observed relative frequencies, corresponding to $\capfracc~\capCI$\% of the magnitude bins in this data set. Hollow dots represents observations outside the bars.}

\def\captitle{Southern California 2015.01.01--2016.01.01, containing {\nrevents} events.}
\def\caplhat{\hat{\lambda}=2.077}
\def\capmc{1.0}
\def\caplmlhat{1.925}
\def\capfrac{70/76=0.921}
\def\capfraca{70}
\def\capfracb{76}
\def\capfracc{92.1}
\def\capCI{(83.8, 96.3)}
\def\nrevents{15046}

\begin{figure}
\begin{center}
\includegraphics[trim={0 0.0cm 0cm 0},clip,scale=1]{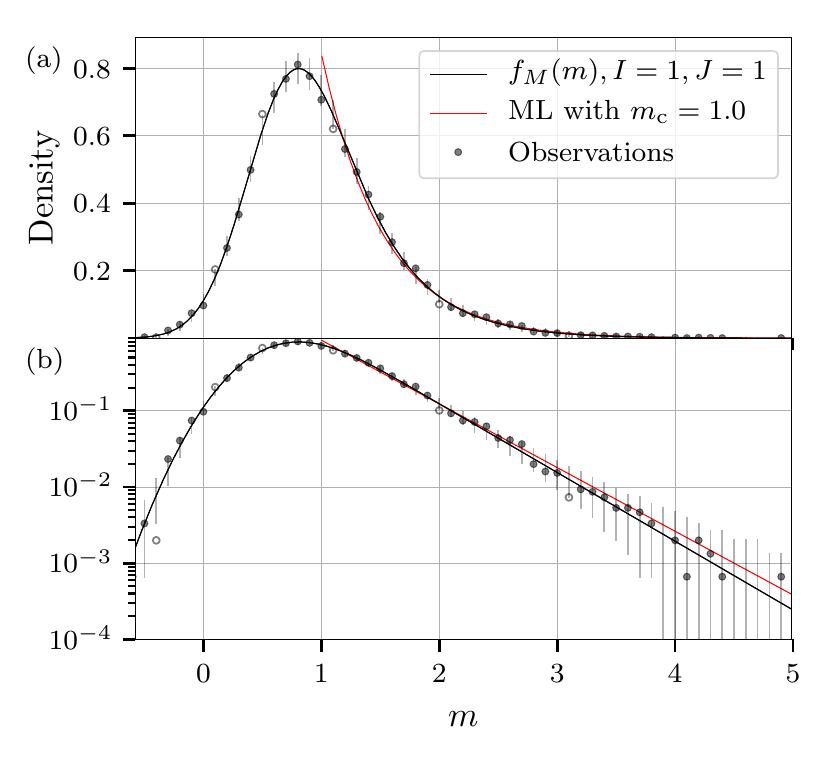}\vspace{-0.5cm}
\caption{\mycapalt}
\label{fig:dataset_1}
\end{center}
\end{figure}

%

The same conclusion can be drawn about the data set in Fig.~\ref{fig:dataset_8}, which shows multi-modal characteristics. In this case, a model with a single term is not flexible enough. Our fitted model for this data set has order $I=2,J=2$. 

\def\captitle{Southern California 2011.06.01--2012.01.01, including {\nrevents} events.}
\def\caplhat{\min_j\hat{\lambda}_j=2.196}
\def\capmc{2.1}
\def\caplmlhat{2.050}
\def\capfrac{80/84=0.952}
\def\capCI{(88.4, 98.1)}
\def\capfraca{80}
\def\capfracb{84}
\def\capfracc{95.2}
\def\nrevents{17879}

\begin{figure}
\begin{center}
\includegraphics[scale=1]{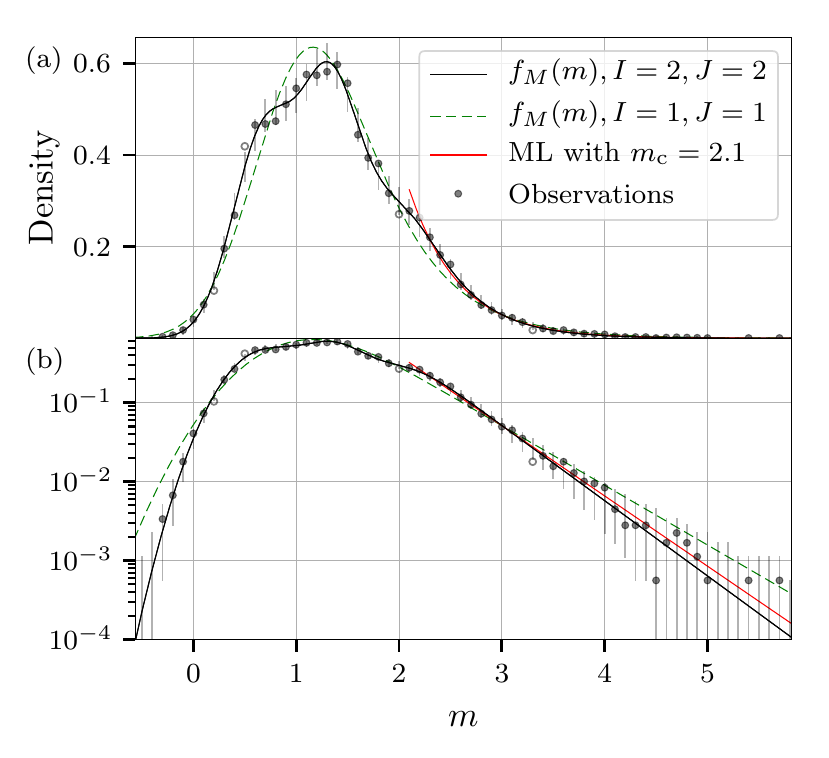}\vspace{-0.5cm}
\caption{\mycapalt}
\label{fig:dataset_8}
\end{center}
\end{figure}


The results in Fig.~\ref{fig:dataset_9} are based on 10 years of data between 1990--2000 and include the $m=7.3$ Landers and the $m=7.1$ Hector mine earthquakes \cite{KWH10}. As expected, the BIC selects a much larger ($I=5,J=2$) model compared with the earlier data sets from the same region, as more data in this case will relieve more details. Also, the effects of non-stationary conditions may be larger over a longer time period and contribute to additional terms in the mixtures. 

\def\captitle{Southern California 1990.01.01--2000.01.01, including {\nrevents} events.}
\def\caplhat{\min_j\hat{\lambda}_j=2.174}
\def\capmc{1.9}
\def\caplmlhat{2.182}
\def\capfrac{87/94=0.926}
\def\capCI{(85.4, 96.3)}
\def\capfraca{87}
\def\capfracb{94}
\def\capfracc{92.6}
\def\nrevents{205555}

\begin{figure}
\begin{center}
\includegraphics[scale=1]{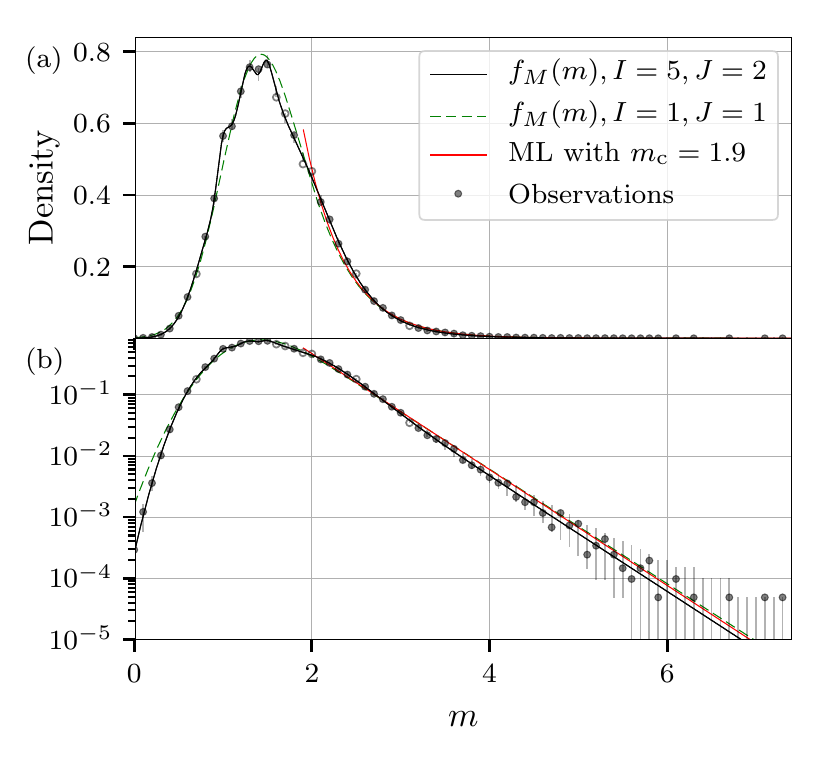}\vspace{-0.5cm}
\caption{\mycapalt}
\label{fig:dataset_9}
\end{center}
\end{figure}


It is worth noticing the slight change in characteristics seen in Fig.~\ref{fig:dataset_9}(b) prior to and after $m=3$, and such a change in slope \cite{Zha17} is captured with the generalization of the GR-law in (\ref{eq:g(m)}) and $J=2$. Although not as prominent as the changes seen for the mining induced data in Fig.~\ref{fig:dataset_6}(b), an environment known to have several characteristics, the additional terms in (\ref{eq:g(m)}) are essential to describe this type of behaviour, as  the detection probability (\ref{eq:P(R<=r_m)}) is by definition monotonically increasing in $m$ and bounded between zero and one.

\def\captitle{LKAB's iron-ore mine in Malmberget, orebody Dennewitz, 2010--2012, including {\nrevents} events.}
\def\caplhat{\min_j\hat{\lambda}_j=2.389}
\def\capmc{-1.4}
\def\caplmlhat{2.474}
\def\capfrac{59/61=0.967}
\def\capCI{(88.8, 99.0)}
\def\capfraca{59}
\def\capfracb{61}
\def\capfracc{96.7}
\def\nrevents{28623}

\begin{figure}
\begin{center}
\includegraphics[scale=1]{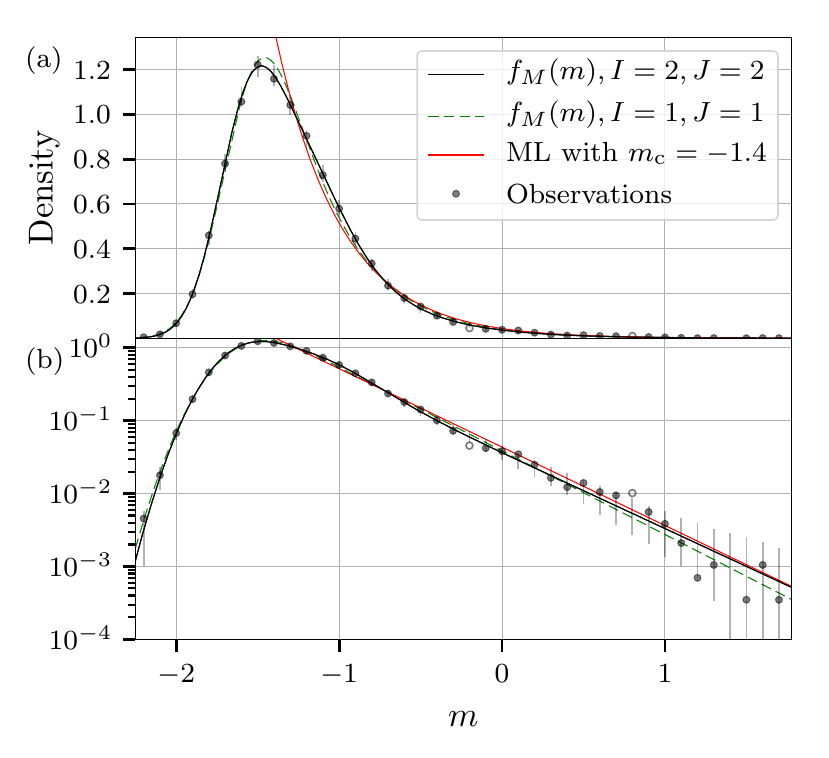}\vspace{-0.5cm}
\caption{\mycapalt}
\label{fig:dataset_6}
\end{center}
\end{figure}

%

Fig.~\ref{fig:dataset_4} concerns the time period 1993--2015 in Japan. No drastic changes or multi-modal characteristics can be seen during this period and there is only a gradual increase in network sensitivity \cite{OKHO04}. Here the BIC selects a model with $I=2$ and $J=1$ to describe the observations during the entire period. 

\def\captitle{Japan 1993--2015, including {\nrevents} events.}
\def\caplhat{\hat{\lambda}=2.133}
\def\capmc{4.0}
\def\caplmlhat{1.981}
\def\capfrac{74/77=0.961}
\def\capCI{(89.2, 98.6)}
\def\capfraca{74}
\def\capfracb{77}
\def\capfracc{96.1}
\def\nrevents{32706}

\begin{figure}
\begin{center}
\includegraphics[scale=1]{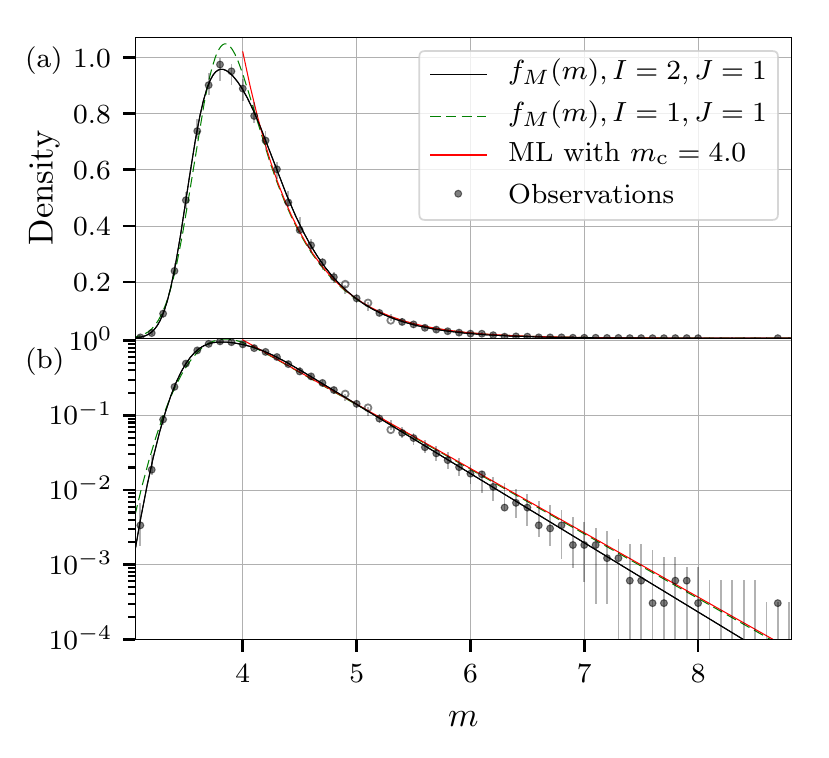}\vspace{-0.5cm}
\caption{\mycapalt}
\label{fig:dataset_4}
\end{center}
\end{figure}

To trust inference based on the model, the replicated data generated using the proposed distribution should resemble the actual observations \cite{GCSR04}. We note that the vertical 95\% bars cover 95\% of the observed relative frequencies in each data set (detailed numbers including confidence intervals are found in figure captions) and that the actual observations look plausible under the proposed distribution \cite{GCSR04}. No systematic discrepancies between the proposed distribution and the observations in Figs.~\ref{fig:dataset_1}--\ref{fig:dataset_4} are observed. 

\begin{figure}
\begin{center}
\includegraphics[scale=1]{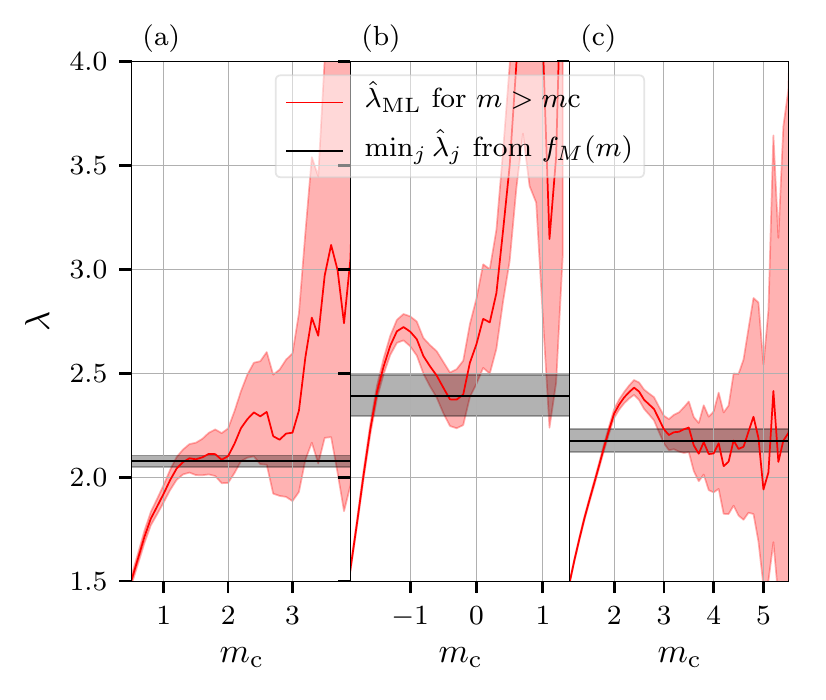}\vspace{-0.5cm}
\caption{{Estimation performance.} Mean and 95\% uncertainties (2.5 and 97.5 percentiles) of $\hat{\lambda}$ using bootstrapping. (a), (b) and (c) represent the least complex model in Fig.~\ref{fig:dataset_1}, the medium size model in Fig.~\ref{fig:dataset_6} and the most complex model in Fig.~\ref{fig:dataset_9}, respectively. The black horizontal line is $\hat{\lambda}$ (or $\min_j\hat{\lambda}_j$ if $J>1$) from $f_M(m)$ with fixed detection probability and is comparable to the red curve that is $\hat{\lambda}_\mathrm{ML}$ at different fixed $m_\mathrm{c}$.}
\label{fig:bs_dataset_19}
\end{center}
\end{figure}

In addition to a description of the detection probability, the model returns a better estimate of the $b$-value in terms of bias and uncertainty for the studied data sets when compared to ML estimation based on events with magnitude $m>m_\mathrm{c}$. In Fig.~\ref{fig:bs_dataset_19} we compare the performances with a fixed detection probability and fixed $m_\mathrm{c}$ to avoid the dependency on a particular method of determining $m_\mathrm{c}$. The left, middle, and right panel in Fig.~\ref{fig:bs_dataset_19} show the estimation performance with the least complex model in Fig.~\ref{fig:dataset_1}, the medium size model in Fig.~\ref{fig:dataset_6}, and the most complex model in Fig.~\ref{fig:dataset_9}, respectively. The results in Fig.~\ref{fig:bs_dataset_19} also affirm the unnecessary trade off between bias and uncertainty at different $m_\mathrm{c}$ briefly mentioned in the first paragraph and illustrates the difficulties in determining a threshold. 

At $m_\mathrm{c}=1.3$, no visible decline in the detection rate is observed in Fig.~\ref{fig:dataset_1} or in Fig.~\ref{fig:bs_dataset_19}(a), and the estimation uncertainty is significantly lower using the proposed model to estimate $\lambda$. The change in $b$-value around $m=3$ in Fig.~\ref{fig:dataset_9} is also visible in Fig.~\ref{fig:bs_dataset_19}(c) as an overestimation between $2<m_\mathrm{c}<3$. As $m_\mathrm{c}$ increases beyond 3, $\hat{\lambda}_\mathrm{ML}$ coincides with $\hat{\lambda}$ from $f_M(m)$ at the cost of higher uncertainty. A similar overestimation is seen for the mining induced data in Fig.~\ref{fig:bs_dataset_19}(b) and a change in $b$-value around $m=-0.5$ is visible in Fig.~\ref{fig:dataset_6}. Compared with 
discarding data below $m_\mathrm{c}$, the detection probability in (\ref{eq:f_M(m)})  increases the estimation performance as information from all events are used to make inference.

\section{Summary}
We make the following conclusions: (i) Our model well describes the distribution of observed magnitudes for the studied data sets. (ii) The model gives richer inference and significantly improved $b$-value estimates in terms of bias and uncertainty. (iii) Without truncation, selection or pre-processing of the observations, the model can describe spatio-temporal variations in the seismic environment (see Eq. (\ref{eq:g(m)})) as well as provide valuable information about the sensor network's limitations (see Eq. (\ref{eq:P(R<=r_m)})). 
(iv) Standard statistical inferential methods are directly utilizable as (\ref{eq:f_M(m)}) is a proper PDF of all the available observations.

\section{Data and Resources}
The data sets from the Southern California Seismic Network can be accessed at \url{https://service.scedc.caltech.edu/eq-catalogs/date_mag_loc.php}. The data set from the National Research Institute for Earth Science and Disaster Prevention in Japan can be accessed at \url{http://www.fnet.bosai.go.jp/event/search.php?LANG=en}. For SCSN and NIED data, we used the default settings for geometrical restrictions. These web sites were last visited in August 2017. The mining induced data in Fig.~\ref{fig:dataset_6} is available from the authors upon request. 

\section*{Acknowledgments}
The authors wish to express their gratitude to the  referees for many valuable comments and suggestions. Financial support from Luossavaara-Kiirunavaara AB (LKAB) is  gratefully acknowledged.

\bibliographystyle{IEEEtran}
\bibliography{ieee_grsl_arxiv}

\end{document}